# Fracture energy of 6H-SiC at the microscale: effects of testing geometry and notch preparation

*Zhuoqi Lucas Li [a,*], Siyang Wang [a], Ao Li [a], James O. Douglas [a], Florian Bouville [a], Oriol Gavalda-Diaz [a], Katharina Tinka Marquardt [a,b,c], Finn Giuliani [a]*

[a] *Department of Materials, Imperial College London, London SW7 2AZ, UK*

[b] *Department of Materials, University of Oxford, Oxford OX2 6NN, UK*

[c] *The Zero Institute, Holywell House, Osney Mead, Oxford OX2 0ES, UK*

*Corresponding author: Zhuoqi Lucas Li (zhuoqi.li17@imperial.ac.uk)*

## Abstract

Micromechanical testing enables small-scale fracture energy measurements, but values depend strongly on geometry and specimen preparation. Here, the fracture energy of the single-crystal 6H-SiC $\{10\bar{1}0\}$ plane was measured using microscale double cantilever beam (DCB) and single cantilever beam (SCB) geometries. DCBs showed stable crack growth under displacement control and obtained 7.5 ± 0.3 J/m$^2$. In contrast, SCBs notched by a Ga focused ion beam gave fracture energies over twice this value, indicating Ga implantation and near-notch residual stresses. Increasing the final notching current increased the measured fracture energy further. Although near-cryogenic notching limited ion-beam-induced damage, it did not reconcile SCB-derived values with DCB test results. Vacuum annealing substantially lowered the fracture energy and brought SCB results into close agreement with DCB measurements, whereas annealing in argon was less effective. Our findings highlight the importance of careful sample preparation and testing geometry selection for reliable fracture property measurement in ceramic materials.

# 1. Introduction

Micromechanical testing has become an increasingly popular approach for probing fracture and deformation behaviours in materials at small length scales [1]. Specifically, fracture energy and fracture toughness measured at the microscale provide critical insight into how local crystallography and interfaces govern bulk fracture behaviour of brittle solids [2]. To date, several techniques, including nanoindentation [3-7], micropillar splitting [8-12], and clamped beam bending [13-16], have been adopted for this application. Among them, double cantilever beams (DCBs) and single cantilever beams (SCBs) are two of the popular geometries for microscale fracture testing [17-31]. DCB and SCB configurations are compatible with focused ion beam (FIB) fabrication, enabling precise placement of notches and crack fronts and allowing well-defined geometries suitable for straightforward analysis [29, 30].

The microscale DCB-based configuration was first proposed by Liu et al. [32]. Although the associated analytical solutions are well established, the flat-punch-loaded DCB requires careful control of sample dimensions and friction conditions to avoid plasticity and to enable meaningful fracture toughness measurements [13, 32]. Later, Sernicola et al. introduced a wedge-tip-loaded DCB configuration operated under displacement control, realised with a piezoelectric actuator, within an scanning electron microscope (SEM), as shown in **Figure 1(a)** [29]. As the wedge advances into the central trough, the two cantilever arms release stored elastic energy to drive crack initiation and propagation. This approach provides stable crack growth and allows multiple fracture energy values at various stages of crack progression to be obtained during a single experiment. The benefit of obtaining multiple fracture-energy values arises from the fact that ion milling can introduce chemical and physical damage in the near-surface region, as discussed in detail by Borasi et al. [33]. Stable crack growth allows potentially inaccurate measurements near the initial notch to be discarded, so that only data recorded after the crack has propagated away from the FIB-affected region are analysed. However, the method requires precise alignment of the wedge tip to ensure symmetric loading, and high-resolution imaging throughout the test is essential for accurate crack-tip tracking and fracture energy determination. In addition, reliable data analysis requires a long, straight crack, making this method less efficient for testing short and complex interfaces [2].

FIB-fabricated microscale SCBs with straight-through notches were first introduced by Di Maio and Roberts (**Figure 1(b)**) [30]. The use of low-current ion beams enables the milling of a sharp notch near the clamped end, promoting crack initiation when the load is applied. Their compatibility with a wide range of indenter geometries, together with a simple loading setup, makes SCB tests straightforward to implement [13, 31, 34]. Furthermore, fracture toughness can be determined using only the peak load and specimen dimensions, enabling both in situ and ex situ testing [30]. Despite these advantages, SCBs are particularly sensitive to FIB-induced damage. With standard room temperature FIB, Ga ion implantation produces compressive residual stresses up to 15 GPa within a 20 nm region ahead of the notch tip in $Al_2O_3$, artificially elevating the measured toughness [35]. The compressive residual stresses responsible for the elevated fracture energy originate from the constrained lattice expansion around the crack tip due to interstitial implantation during milling [35]. Alternative milling ions such as Xe and He have been explored and shown to yield slightly lower toughness values [36], while Ne ion milling may introduce gas-bubble defects that complicate data interpretation [37]; similarly, Xe FIB has also been shown to result in bubble formation [38]. Although Xe and He ion milling are shown to be promising in suppressing preparation-induced defects, Ga ion FIB systems remain the most widely available and commonly used. Post-notching heat treatment of SCBs offers a route to relieve residual stresses, for instance, Norton et al. demonstrated that annealing Ga-milled $Al_2O_3$ SCBs at 1000-1200 °C substantially reduces the apparent stress intensity factor and thus improves the reliability of the toughness measurements [35].

An alternative mitigation strategy is to reduce damage formation and redistribution *during* the milling process. Cryogenic FIB (cryo-FIB) milling is therefore attractive because lowering the specimen temperature can suppress thermally activated defect evolution and limit the mobility of implanted species during irradiation and immediately afterward [39]. In titanium alloys, cryo-FIB has been shown to strongly reduce preparation artefacts associated with hydrogen uptake and hydride formation [39], while in aluminium alloys it can limit the redistribution and segregation of implanted Ga into grain boundaries compared with room-temperature milling [40]. Although cryo-FIB has been reported to reduce preparation-induced damage in environmentally sensitive materials and to restrict Ga redistribution in certain systems, its value for microscale fracture testing for brittle materials has not yet been clearly established. This issue is particularly important for SCBs with straight-through notches because the crack initiates directly from the FIB-machined notch. As a result, even a very shallow region affected by ion implantation at the notch root may strongly influence the measured fracture properties. If effective, cryogenic Ga-FIB notching would provide a practical mitigation strategy that avoids the need for specialised alternative ion-source instruments or additional post-fabrication thermal processing. Therefore, a direct evaluation of whether cryogenic Ga-FIB notching can improve the reliability of SCB-based fracture energy and toughness measurements is necessary.

Given the complicated effects of test geometries and FIB-induced damage on fracture property measurements at small scale, a thorough investigation, using a single reference material, into how each individual variable such as FIB current, milling temperature, and post-mortem annealing temperature, quantitatively modulate the property data obtained, is essential for establishing a clear understanding of any potential inaccuracies involved in a specific test procedure, as well as the best practices in designing test strategies for assessing small scale fracture behaviour of hard materials. This study therefore presents a systematic comparison of the fracture energy measured experimentally using two widely employed micromechanical geometries, DCBs and SCBs, fabricated on the prismatic plane of single-crystal 6H-SiC. In addition to evaluating intrinsic geometric differences, particular emphasis is placed on improving the reliability of SCB-based measurements, which are highly sensitive to Ga-induced residual stresses and implantation damage. To this end, we investigate several notch-fabrication and post-processing strategies, including the use of different ion-beam currents, near-cryogenic notching, and a range of annealing conditions at different environments. Through these comparative and optimisation efforts, this work establishes how testing geometry and notch preparation in SCBs influence microscale fracture measurements and provides guidelines for obtaining accurate and reproducible fracture energy and toughness values at small length scales.

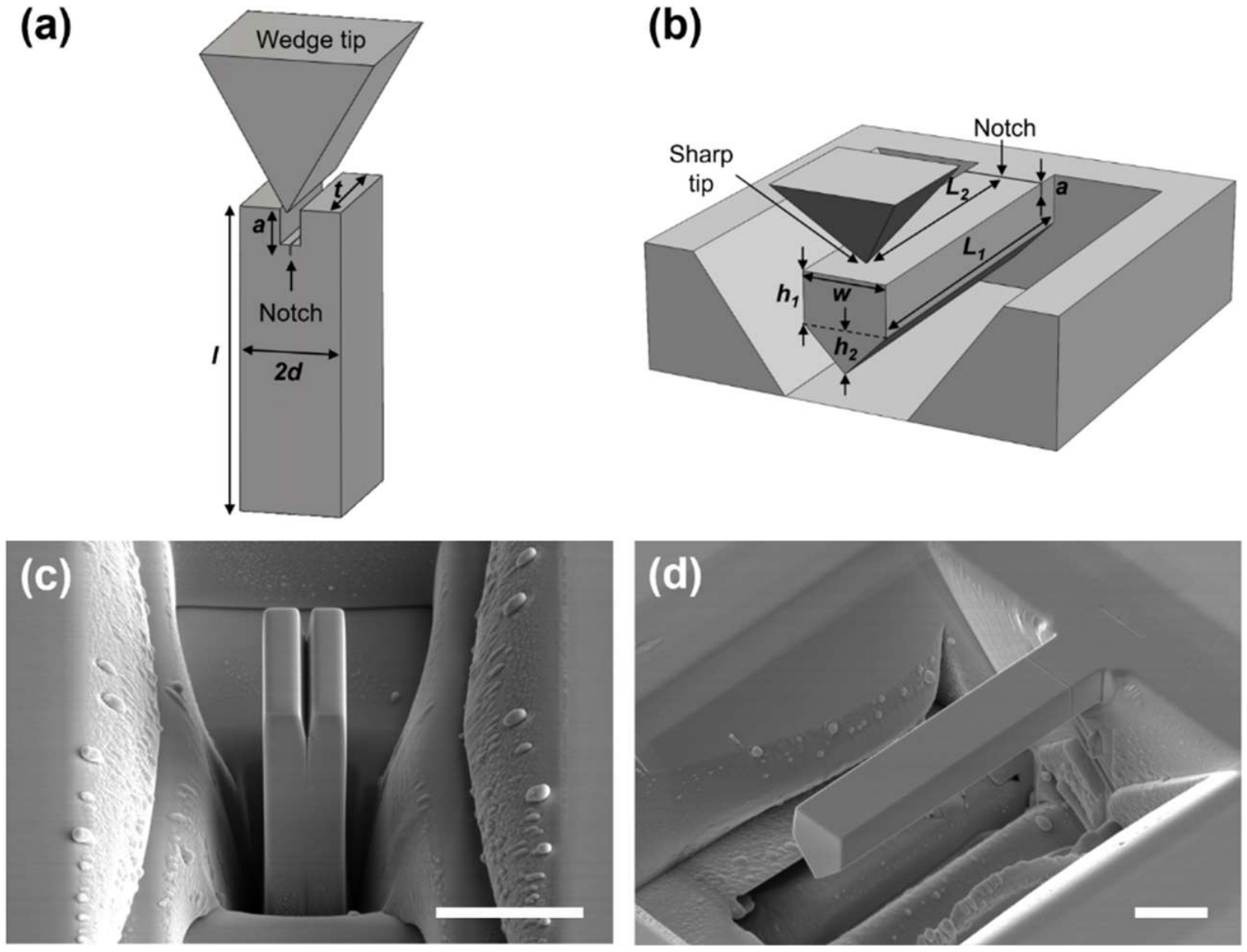

**Figure 1.** Geometries employed for micromechanical fracture toughness/energy measurement. (a) Schematic and (c) electron micrograph of a DCB. (b) Schematic and (d) electron micrograph of an SCB with a straight-through notch. For the DCB geometry, *l* denotes the beam height, *2d* the total cantilever arm width, *a* the crack length, and *t* the beam thickness. For the SCB geometry, *w* denotes the beam width, *a* the crack length, $L_1$ the total beam length, $L_2$ the distance between the notch and the loading point, and $h_1$ and $h_2$ the corresponding heights of the upper and lower parts of the beam, respectively. Scale bars: 5 μm.

## 2. Materials and Methods

### 2.1. Materials

Single-crystal 6H-SiC wafers (5×5×0.33 mm) were purchased from MTI Corporation, with a surface out-of-plane orientation of <0001>.

### 2.2. DCB fabrication

DCBs were fabricated using a Thermo Fisher Scientific Scios 2 or Helios 5 CX DualBeam $Ga^+$ FIB-SEM with a 30 kV ion accelerating voltage. Large ion beam currents (~20 nA and ~9 nA) were employed to create a trench around the initial pillar. The pillars were then cleaned with 1 nA FIB; the sample was over-tilted and under-tilted by 1° to reduce tapering, ensuring a constant cross-sectional area along the entire height. A trough was subsequently cut on the top surface using a smaller beam current of ~0.3 nA. Finally, a notch was cut using a line pattern and a ~10 pA beam current. Typical dimensions for the DCBs used in this work are: $10 < l < 15$ µm, $2d \sim 2.6$ µm, $a \sim 1.5$ µm, and $t \sim 5$ µm (**Figure 1(a)**).

## 2.3. SCB fabrication

SCBs were fabricated using the same FIB-SEM systems as employed for DCB fabrication. A trench was initially milled with a ~20 nA ion beam. The sample was then tilted to 45° with respect to the ion beam, and a ~9 nA beam current was used to cut through the beam structure. Subsequently, the sample was rotated 180° to undercut the opposite side. The beam end and two vertical sidewalls were then cleaned using a ~1 nA ion beam. The same beam current was subsequently used to clean the undercut surfaces; during this step, the sample was over-tilted by 1° to compensate for beam divergence and maintain the desired geometry. Finally, a straight-through notch was milled. Typical dimensions for the SCBs used in this work are $w \sim 5$ µm, $a \sim 0.2$ to 0.4 µm, $L_1 \sim 27$ µm, $L_2 \sim 22$ µm, $h_1 \sim 3$ µm, and $h_2 \sim 2.5$ µm. For SCBs, the notch was milled at approximately 3 µm from the clamped end (**Figure 1(b)**).

## 2.4. In situ micromechanical testing

Both the DCB and SCB fracture experiments were conducted using an Alemnis nanoindenter in an FEI Quanta 650 SEM. For the DCB tests, a 70° diamond wedge indenter (Synton-MDP AG, Switzerland) with a nominal tip length of 10 µm was used. The wedge tip was aligned with the DCBs using a rotational stage to achieve precise positioning between the notch and the wedge. During the test, a tip displacement rate of 1 nm/s was employed. Images were acquired every second during the test for subsequent analysis. For the SCB tests, a cube-corner indenter with a sharp tip (Synton-MDP AG, Switzerland) was used. The cantilever was loaded near its free end at a constant displacement rate of 5 nm/s. An image was acquired at the start of the test to measure the distance from the notch to the loading point ($L_2$ in **Figure 1(b)**). After the test, the depth of the initial FIB-milled notch was measured from the fracture surface.

## 2.5. Determination of fracture energy and fracture toughness

### 2.5.1. DCB geometries

For DCB testing geometries, the critical elastic energy release rate/fracture energy, $G_c$ , was calculated using **Equation 1** [2, 27, 28, 41, 42]:

$$G_c = \frac{3Ed^3\delta^2}{4(a+\chi d)^4}\left[1 + (1+v)(\frac{d}{(a+\chi d)})^2\right]$$

**Equation 1**

where $E$, $d$, $\delta$, $a$ and $v$ are the elastic modulus, cantilever arm width, arm lateral displacement, crack length and Poisson's ratio respectively. Here, $\chi$ is a correction factor that addresses crack tip rotation and was approximated as 0.67 [41, 43]. During the test, the crack length $a$ was measured from the edge of the top surface, including both the initial trough length and the notch length introduced by FIB. The elastic modulus for the orientation parallel to the prismatic $\{10\bar{1}0\}$ planes was used (553 GPa) [44]. As the crack evolves, a >> d and **Equation 1** reduces to **Equation 2**:

$$G_c = \frac{3Ed^3\delta^2}{4(a+\chi d)^4}$$

**Equation 2**

### 2.5.2. SCB geometries

The critical stress intensity factor/fracture toughness, $K_{Ic}$, was calculated from single-notched SCB geometries with **Equation 3**, as described by Chen et al. [34]:

$$K_{Ic} = \sigma_f\sqrt{\pi a} \cdot F(\frac{a}{h_1+h_2})$$

**Equation 3**

where $F(\frac{a}{h_1+h_2})$ is a dimensionless shape factor that depends on $a$, the crack length and $h_1$ and $h_2$, the upper and lower height of the beam (**Figure 1(b)**). $\sigma_f$ is the fracture stress and is given by **Equation 4**:

$$\sigma_f = \frac{PL_2y}{I}$$

**Equation 4**

where $P$ is the load of failure, $L_2$ is the distance between the notch and the loading point. Here, $y$ and $I$ represent the vertical distance between the upper surface and the neutral plane, and the moment of inertia of the pentagonal cross-section, respectively. They are expressed as in **Equation 5** and **Equation 6**:

$$y = \frac{6{h_1}^2 + 3h_1h_2 + {h_2}^2}{12h_1 + 3h_2}$$

**Equation 5**

$$I = \frac{wh_1^{\ 3}}{12} + (y - \frac{h_1}{2})^2 h_1 w + \frac{wh_2^{\ 3}}{36} + (\frac{h_2}{3} + h_1 - y)^2 \frac{wh_2}{2}$$

**Equation 6**

where $w$ is the beam width. In this work, the beam geometry-dependent shape factor $F(\frac{a}{h_1+h_2})$ was determined by finite element analysis (FEA) in Abaqus (Abaqus/CAE 2022, Dassault Systèmes) [34, 45]. A beam with the typical dimensions shown in **Figure 1(b)** was first modelled. The material was modelled as elastic and anisotropic. The elastic constants for 6H-SiC were $C_{11}$ = 501 GPa, $C_{33}$ = 553 GPa, $C_{44}$ = 163 GPa, $C_{12}$ = 111 GPa, and $C_{13}$ = 52 GPa [44]. An encastre boundary condition was applied at the clamped end of the beam. A series of notches of various lengths were introduced, and the beam was loaded with a concentrated point force located a distance $L_2$ from the crack. The stress intensity factors at the crack tip were obtained and averaged using the contour integral method. The shape factors corresponding to different notch lengths were then calculated. Finally, an $F(\frac{a}{h_1+h_2})$ versus $\frac{a}{h_1+h_2}$ ($0.0375 \le \frac{a}{h_1+h_2} \le 0.4$) plot was created and the data points were fitted to a 4th order polynomial function to obtain a general shape function applicable to all SCBs used in this study, as shown in **Equation 7**:

$$F\left(\frac{a}{h_1+h_2}\right) = 1.35 - 2.01\left(\frac{a}{h_1+h_2}\right) + 13.68\left(\frac{a}{h_1+h_2}\right)^2 - 29.14\left(\frac{a}{h_1+h_2}\right)^3 + 33.94\left(\frac{a}{h_1+h_2}\right)^4$$

**Equation 7**

For SCB results, the calculated fracture toughness was converted to fracture energy using the relationship in **Equation 8**:

$$G_C = \frac{K_{Ic}^{\ 2}}{E_\perp}$$

**Equation 8**

where $E_\perp$ is the elastic modulus perpendicular to the splitting plane. An elastic modulus of 501 GPa was used for the fracture energy calculation of the prismatic $\{10\bar{1}0\}$ planes [44]. Given the small value of the Poisson's ratio ($\nu_{13} \approx 0.07$), the difference between plane stress and plane strain conditions is negligible [29, 44].

## 2.6. Notching and annealing conditions

To examine how notch fabrication and post-processing influence the measured fracture energy of the SCB geometry, we systematically varied the notching conditions and applied post-notching annealing treatments (**Table 1**). All FIB notching was conducted with 30 kV ion accelerating voltage. First, the final notching current was increased from 10 pA to 30 pA at room temperature (RT10 and RT30) to evaluate the effect of beam current on the measured fracture energy. Next, final notching was performed at a near-cryogenic temperature (-127 °C) at 10 pA, 30 pA, and 50 pA (Cryo10, Cryo30, and Cryo50) to test the hypothesis that reducing the specimen temperature during the final notching step would suppress defect mobility and Ga redistribution near the notch root, thereby improving the reliability of the resulting fracture energy values. A Thermo Fisher Scientific CryoMAT stage was used, which cools the stage by flowing $N_2$ gas passed through liquid $N_2$ at 2.1 bar via a heat exchanger. The temperature was controlled through a combination of $N_2$ gas flow regulation and active heating. During notching, a fixed milling depth setting of 2 µm (as calibrated on silicon provided by the manufacturer) was used for all ion beam currents and notching conditions. Because the beam current can vary due to factors such as aperture wear, the actual current during notch milling was verified using inline Faraday cup measurements to be close to the selected currents. A line pattern was created adjacent to the SCB trench region and used to fine-tune focus and stigmation to ensure optimal and consistent notch sharpness (**Figure 1(b)**).

Finally, post-notching annealing was carried out at 600 °C and 800 °C in vacuum (Va600 and Va800), and at 800 °C in argon (Ar-10 vol% $H_2$) supplied by BOC (Ar800) with a Thermal Technology LLC 1100-2560-W2 vacuum furnace, to determine whether thermal treatment and annealing atmosphere reduced Ga ion implantation-related effects and yielded more representative fracture energy values. The wafer containing notched SCBs was annealed at 600 °C or 800 °C for 1 hour with a heating and cooling ramp rate of 2 °C/min in either an inert Ar atmosphere (~$10^5$ Pa) or under vacuum ($10^{-3}$ to $10^{-4}$ Pa). The two temperatures were selected relative to the estimated boiling temperature of Ga under the vacuum conditions. Taking 2400 °C as the boiling point at 1 atm, and 256 kJ/mol as the enthalpy of vaporisation [46], and using the Clausius-Clapeyron relation [47], the boiling point of Ga between $10^{-3}$ and $10^{-4}$ Pa is approximately 755 °C and 681 °C, respectively. Accordingly, 600 °C lies below this range and is expected to limit Ga evaporation while still allowing Ga to redistribute and diffuse away from the notch tip, whereas 800 °C lies above it and is expected to promote more complete Ga removal from the material by evaporation. During annealing, a freshly polished hafnium metal piece was placed next to the sample to act as an oxygen getter and reduce potential oxidation of the SiC sample.

## 2.7. Energy-dispersive X-ray spectroscopy (EDS) analysis

After annealing, EDS was used to reveal Ga distribution in and around the FIB-prepared samples. Analyses were performed using a Bruker XFlash® 6|60 detector mounted on an FEI Quanta FEG 650 SEM, operated at 20 kV accelerating voltage with a 30 µm aperture, spot size 5.5, and a 14.9 mm working distance.

**Table 1.** Summary of the notching and post-notching annealing conditions used for all SCB specimens. Sample names reflect the preparation route: RT10 and RT30 correspond to room-temperature notching at 10 pA and 30 pA, respectively. Cryo10, Cryo30 and Cryo50 denote near-cryogenic notching performed at 10 pA, 30 pA, and 50 pA. Va600 and Va800 correspond to specimens annealed in vacuum at 600 °C and 800 °C for 1 h, and Ar800 denotes annealing at 800 °C in Ar for 1h;

| SCB groups | Notching current (pA) | Notching condition | Notching time (s) | Annealing conditions |
|---|---|---|---|---|
| RT10 | 10 | Room temperature | 120 | - |
| RT30 | 30 | Room temperature | 43 | - |
| Cryo10 | 10 | Near-cryogenic (-127 °C) | 120 | - |
| Cryo30 | 30 | Near-cryogenic (-127 °C) | 43 | - |
| Cryo50 | 50 | Near-cryogenic (-127 °C) | 29 | - |
| Va600 | 10 | Room temperature | 120 | 600 °C in vacuum |
| Va800 | 10 | Room temperature | 120 | 800 °C in vacuum |
| Ar800 | 10 | Room temperature | 120 | 800 °C in Ar |

# 3. Results

## 3.1. 6H-SiC fracture energy via DCB geometry

DCB splitting experiments were first performed on the prismatic $\{10\bar{1}0\}$ plane of single-crystal 6H-SiC. As shown in **Figure 2(a)**, after the DCB was loaded, the crack propagated along a straight path and its growth remained stable as the indenter tip travelled deeper. At 540 s, retraction of the indenter tip resulted in crack closure, further indicating stable and reversible opening under displacement control. The post-fracture morphology of one of the DCBs in **Figure 2(b)** revealed a smooth and flat fracture surface, with no evidence of crack branching or deflection across multiple planes before the final rupture of the right arm, at a crack length of ~8 μm.

One key advantage of the DCB geometry is that it enables the tracking of the fracture energy evolution as the crack advances during a single test. **Figure 2(c)** shows the evolution of the calculated fracture energy with crack length for four DCB specimens. Initially elevated fracture energy values gradually decreased and stabilised as the crack advanced. Once the crack length exceeded approximately 5 μm, the measured fracture energy converged across all four beams, resulting in an average fracture energy $G_c$ of 7.5 ± 0.3 J/m$^2$ ($K_{Ic}$ = 1.93 ± 0.04 MPa·m$^{0.5}$).

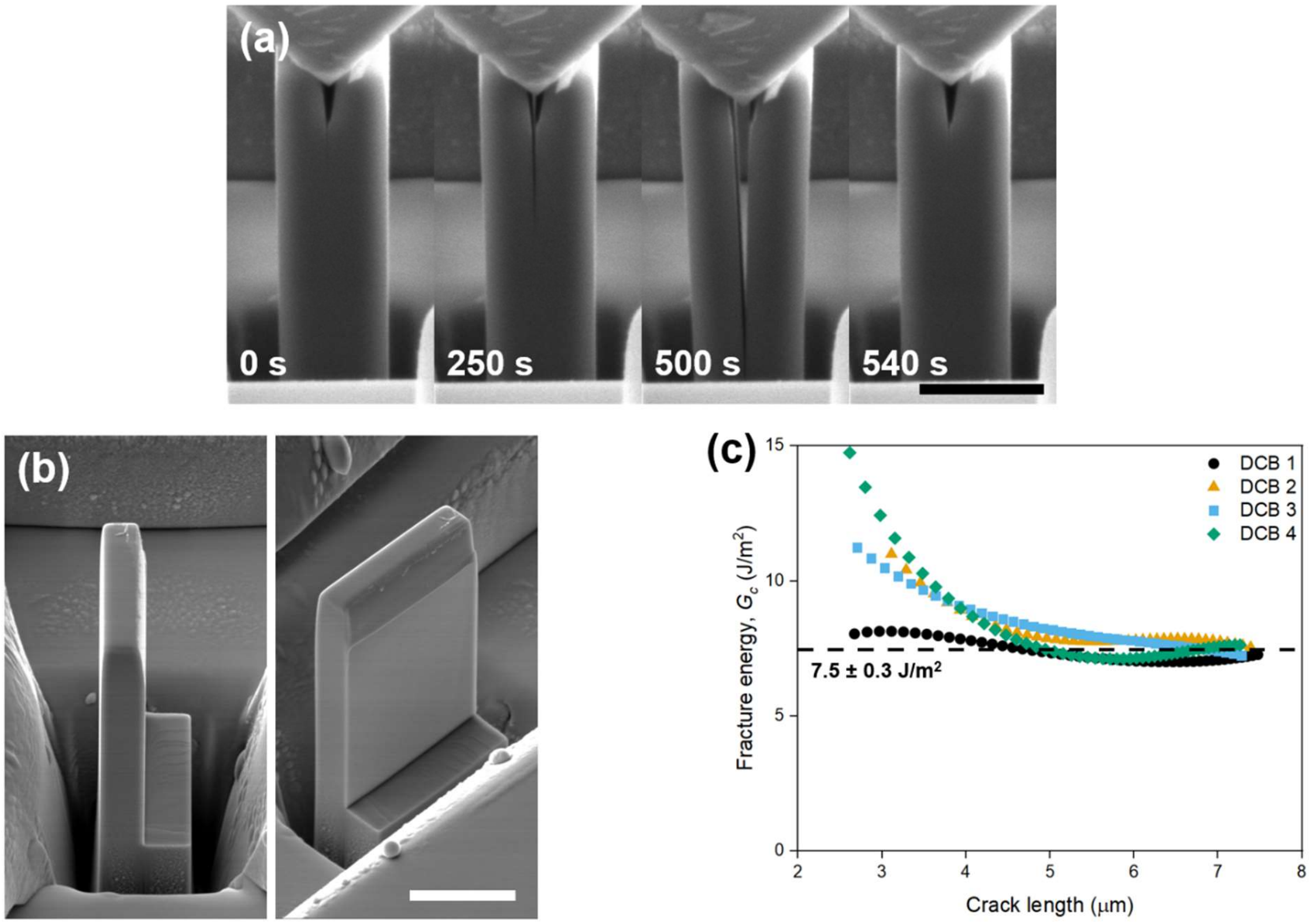

**Figure 2.** (a) The DCB testing process, illustrating slow and stable crack growth along the prismatic $\{10\bar{1}0\}$ plane at different time points. At 540 s, retraction of the indenter tip results in crack closure. (b) Fracture along the prismatic plane of the 6H-SiC single crystal in DCB 4, highlighting the straight and smooth fracture surface created during loading. Scale bars: 3 μm. (c) Fracture energy values measured as the crack propagates. The dashed line indicates the average of all data points beyond a total crack length of 5 μm, after which the calculated fracture energy values stabilise across all four beams.

## 3.2. 6H-SiC fracture energy via SCB geometry

SCBs notched at room temperature using a 10 pA final current (Group RT10 in **Table 1**) exhibited a fracture energy of $G_c = 17.9 \pm 0.4$ J/m$^2$ ($K_{Ic} = 2.99 \pm 0.03$ MPa·m$^{0.5}$), which is more than twice the DCB-derived value (**Figure 3(a) and (b)**). In addition to the elevated fracture energy values, all specimens within this group displayed a distinct horizontal platform along the basal plane of 6H-SiC, near the lower portion of the fracture surface (**Figure 3(c)**), rather than a uniformly smooth surface along the prismatic plane.

### 3.2.1. The effect of ion beam current (RT10 and RT30)

As shown in **Figure 3(a)**, the measured fracture energy for RT30 (30 pA final notching) was $G_c = 22 \pm 3$ J/m$^2$, compared with $G_c = 17.9 \pm 0.4$ J/m$^2$ for RT10 (10 pA final notching). For fracture toughness, the value changed from $K_{Ic} = 2.99 \pm 0.03$ MPa·m$^{0.5}$ (RT10) to $K_{Ic} = 3.3 \pm 0.2$ MPa·m$^{0.5}$ (RT30). In addition, a platform feature similar to that observed for RT10 was evident on the RT30 fracture surfaces (**Figure 3(c)**). Taken together, increasing the final notching current from 10 pA to 30 pA at room temperature yielded higher values of $G_c$ and $K_{Ic}$ and retained the platform feature on the fracture surface.

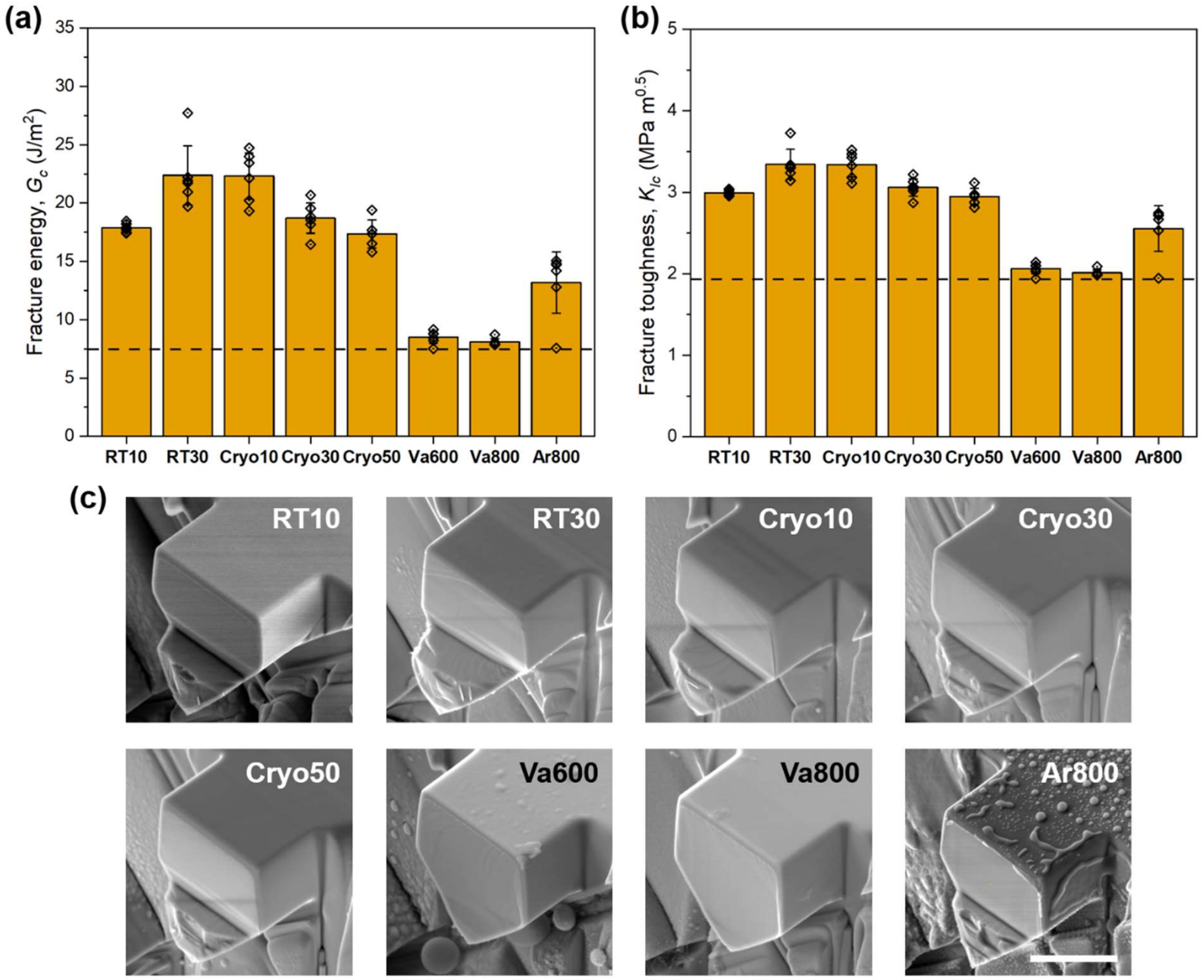

**Figure 3.** (a) Fracture energy, $G_c$, and (b) fracture toughness, $K_{Ic}$, measured for the eight sample groups (RT10, RT30, Cryo10, Cryo30, Cryo50, Va600, Va800, and Ar800). Bar heights indicate the mean value for each group, and open diamonds denote individual measurements. Error bars represent standard deviations. Dashed horizontal lines mark the corresponding values obtained from DCB tests ($G_c$ = 7.5 J/m$^2$ in (a) and $K_{Ic}$ = 1.93 MPa·m$^{0.5}$ in (b)). (c) SEM images of representative fracture surfaces for each sample group. Scale bar: 3 μm.

### 3.2.2. Near-cryogenic FIB notching (Cryo10, Cryo30 and Cryo50)

The fracture energies and fracture toughness obtained from the SCBs notched at the near-cryogenic temperature (Cryo10/30/50) are shown in **Figure 3(a) and (b)**.

At 10 pA (Cryo10), low-temperature notching did not reduce the measured fracture energies relative to the room-temperature group (RT10). Specifically, Cryo10 exhibited $G_c$ = 22 ± 2 J/m$^2$ ($K_{Ic}$ = 3.3 ± 0.2 MPa·m$^{0.5}$), compared with $G_c$ = 17.9 ± 0.4 J/m$^2$ for RT10. Increasing the notching ion beam current to 30 pA and 50 pA led to lower $G_c$ values than Cryo10. Cryo30 showed $G_c$ = 19 ± 1 J/m$^2$ ($K_{Ic}$ = 3.1 ± 0.1 MPa·m$^{0.5}$), while Cryo50 yielded $G_c$ = 17 ± 1 J/m$^2$ ($K_{Ic}$ = 3.0 ± 0.1 MPa·m$^{0.5}$). Despite this small reduction with increasing current, the fracture energies obtained from all low-temperature-notched groups remained well above the DCB values ($G_c$ = 7.5 ± 0.3 J/m$^2$).

Representative fracture surfaces for Cryo10, Cryo30, and Cryo50 groups are shown in **Figure 3(c)**. Similar platform features to those observed from the room temperature-notched specimens retained.

#### 3.2.3. Post-notching annealing (Va600, Va800 and Ar800)

Post-notching annealing was then investigated as a practical route to mitigate the influence of Ga implantation and improve the reliability of SCB-derived fracture properties. The sample surface conditions after annealing are shown in **Figure 4**. Following vacuum annealing at 600 °C for 1 h (Va600), spherical-shaped features were observed in the vicinity of the beam and surrounding trench. Similarly, after vacuum annealing at 800 °C for 1 h (Va800), collapsed shell-like features were observed within the SCB trench. Examples are highlighted by dotted and dashed arrows in **Figure 4(a)** and **(b)**. In contrast, annealing at 800 °C in Ar for 1 h (Ar800) produced widespread particles on the SCB and surrounding surfaces (**Figure 4(c)**). Higher-magnification imaging of the notch region in **Figure 4(d)** further confirmed that particles were also present within the notch.

The corresponding Ga (re-)distribution was further illustrated by the EDS Ga maps overlaid on the SEM images in **Figure 5(a)-(d)**. The as-fabricated SCB specimen showed Ga concentration within the trench. After vacuum annealing at 600 °C for 1 h (Va600), from **Figure 5(b)**, the spherical features observed are confirmed to be Ga-rich, indicating that these features correspond to spherical Ga particles formed during annealing. In contrast, the collapsed shell-like features observed in the Va800 samples (dashed arrows in **Figure 5(c)**) are not Ga-enriched in the EDS map. Consistent with the SEM images, the specimen annealed in Ar (Ar800) showed detectable Ga across the analysed region in **Figure 5(d)**, both within the trench and on the surrounding surfaces outside the trench. Together with the particle-like morphology observed in **Figure 4(c)**, this indicates that the surface features formed on Ar800 specimens after annealing correspond to resolidified Ga particles that had melted at elevated temperature. A notable observation was that Ga was detectable outside the SCB trench for both Va800 and Ar800 specimens. In contrast, for the as-fabricated and Va600 specimens (**Figure 5(a) and (b)**), the Ga signal was largely confined to the SCB trench. **Figure 5(e)** further compares the relative Ga content through the Ga Kα peak intensity for the as-fabricated and annealed specimens. All annealed samples exhibited a lower Ga Kα peak than the as-fabricated sample, indicating an overall decrease in detectable Ga after annealing. Among the annealed groups, vacuum annealing at 800 °C (Va800) showed the lowest Ga peak intensity, likely as a result of Ga evaporation at this temperature under vacuum.

The corresponding fracture energy and toughness values are shown in **Figure 3(a) and (b)**. Both vacuum-annealing treatments resulted in a substantial reduction in $G$ compared to the non-annealed (as-fabricated) SCBs. Specifically, the measured fracture energies were $G_c$ = 8.5 ± 0.5 J/m$^2$ ($K_{Ic}$ = 2.06 ± 0.06 MPa·m$^{0.5}$) for Va600 and $G_c$ = 8.1 ± 0.3 J/m$^2$ ($K_{Ic}$ = 2.01 ± 0.04 MPa·m$^{0.5}$) for Va800, bringing the SCB-derived fracture energy to a level comparable to the DCB value (7.5 ± 0.3 J/m$^2$). In comparison, annealing in Ar at 800 °C (Ar800) produced a more moderate reduction, with $G_c$ = 13 ± 3 J/m$^2$ ($K_{Ic}$ = 2.6 ± 0.3 MPa·m$^{0.5}$), which remains substantially above the DCB results.

Fracture surfaces for the annealed groups are shown in **Figure 3(c)**. After vacuum annealing (Va600 and Va800), the fracture surfaces appeared smooth and planar, and the platform feature observed in other sample groups was absent. In contrast, the Ar-annealed Ar800 specimen showed a clear platform feature and rougher surrounding surfaces due to the Ga particle coverage observed after annealing (**Figure 4(c)**).

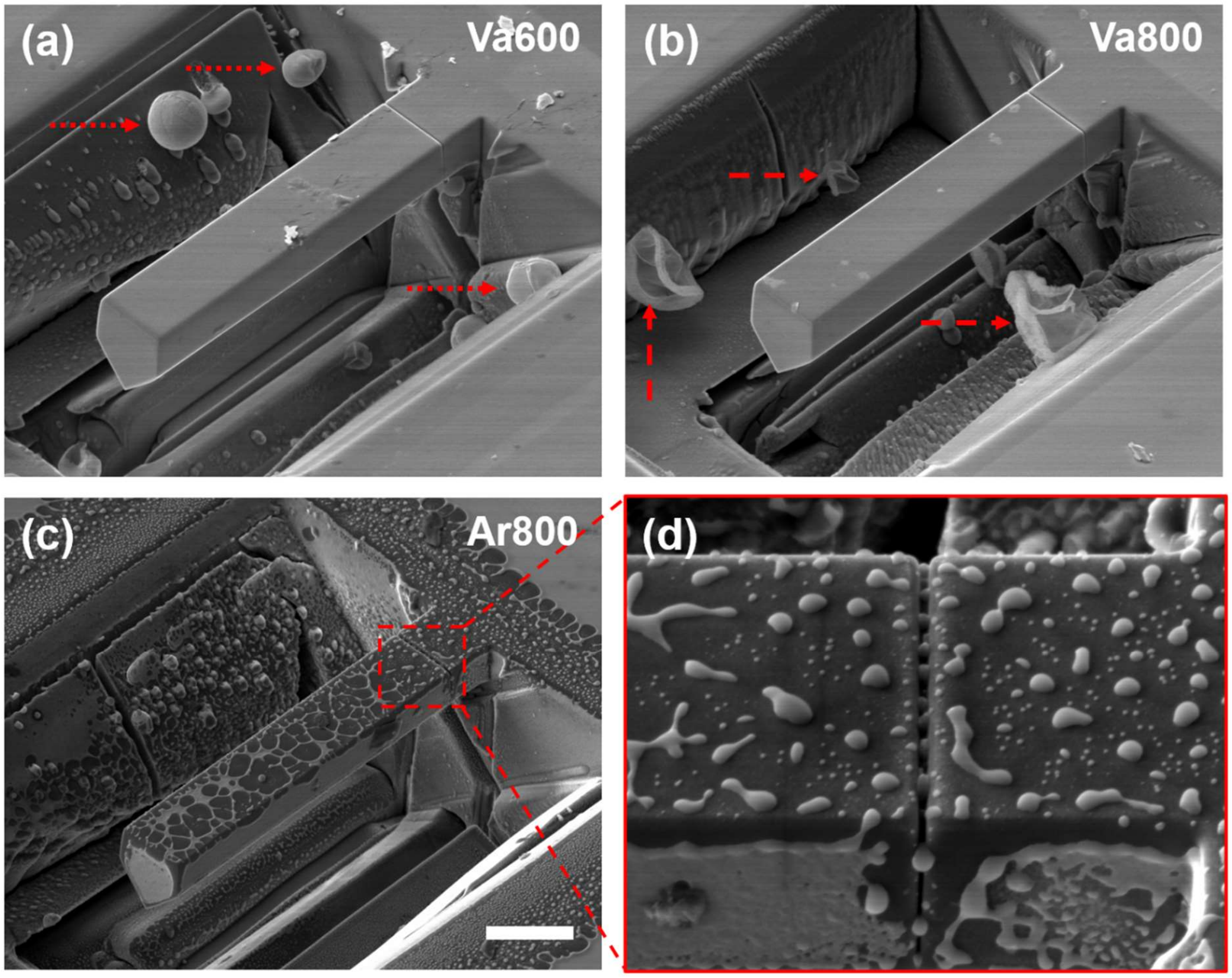

**Figure 4.** SEM images of the SCBs after annealing treatment: (a) 600 °C in vacuum for 1 h (Va600), (b) 800 °C in vacuum for 1 h (Va800) and (c) 800 °C in Ar for 1 h (Ar800). Red arrows in (a) and (b) highlight the representative surface features observed after annealing. Scale bar: 5 µm. (d) Higher-magnification view of the notch region outlined in (c), showing Ga particles within the notch.

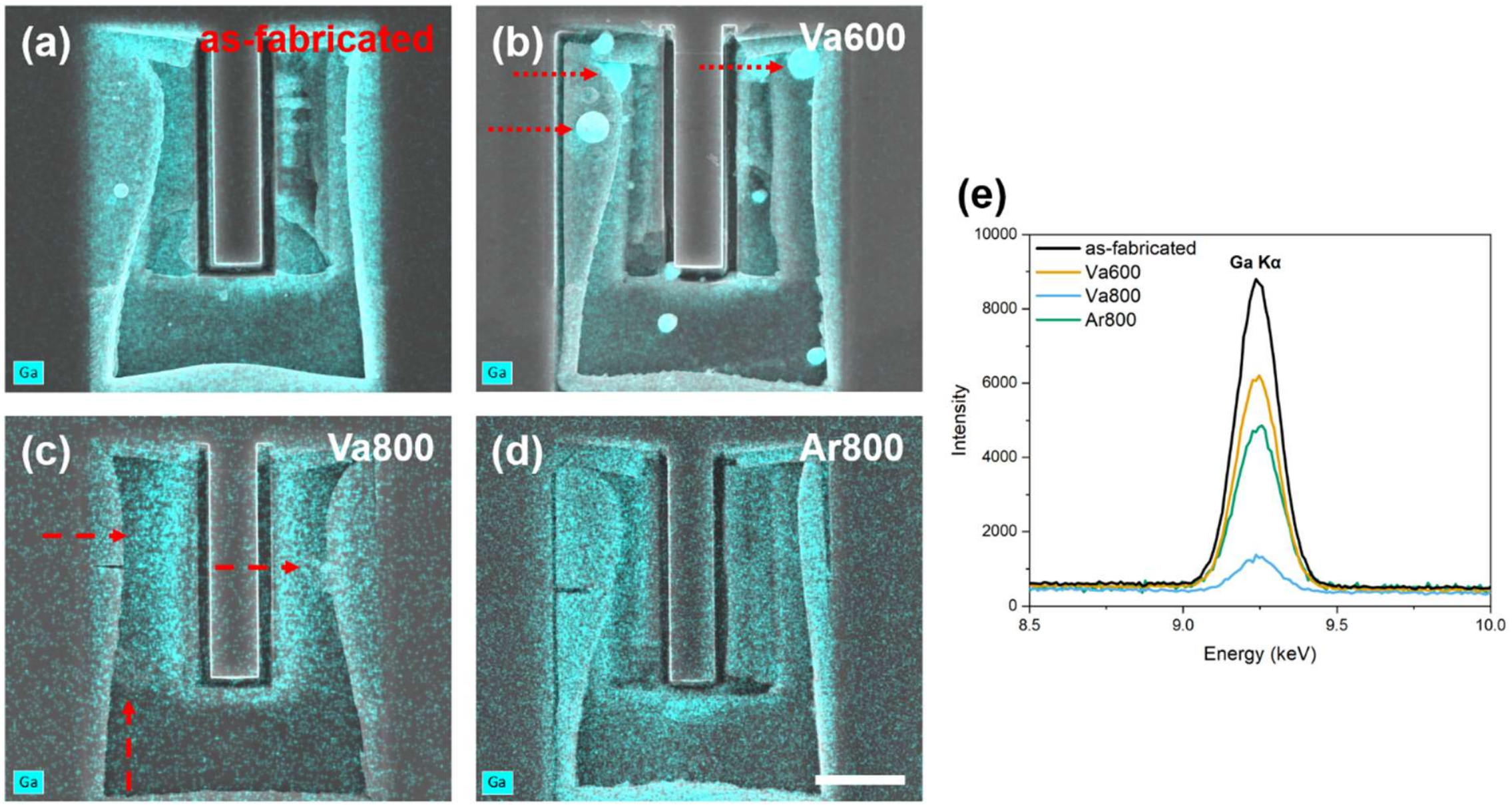

**Figure 5.** EDS Ga maps overlaid on SEM images for the (a) as-fabricated (non-annealed) specimen and specimens annealed for 1 h at (b) 600 °C in vacuum (Va600), (c) 800 °C in vacuum (Va800), and (d) 800 °C in Ar (Ar800). Red arrows highlight the same representative surface features in **Figure 4**. Scale bar: 10 µm. (e) Corresponding EDS spectra in the Ga Kα region (~9.25 keV) comparing relative Ga signal intensity for the as-fabricated and annealed specimens.

### 3.2.4. Platform length and fracture energy

During the examination of the SCB fracture surfaces, specimens exhibiting higher measured fracture energies were consistently associated with a more pronounced step-edge platform feature, whereas specimens with lower $G_c$ values showed little to no platform (**Figure 3**). To confirm this observation, the measured $G_c$ values for all eight sample groups were plotted against the platform lengths that were measured from SEM images (**Figure 6(a)**).

**Figure 6(a)** shows a positive relationship between platform length and fracture energy across the dataset. Specimens in groups Va600 and Va800, which exhibited the lowest calculated fracture energies that were close to the DCB-derived results, showed no platform. In contrast, specimens with elevated fracture energies consistently exhibited larger platform lengths (>0.5 µm). Upon closer examination of the loading process, the platform length was found to correlate with the maximum beam-end deflection (the deflection immediately before fracture), as shown in **Figure 6(b)**. SCBs that reached larger beam-end deflections generally exhibited longer platforms on the fracture surfaces.

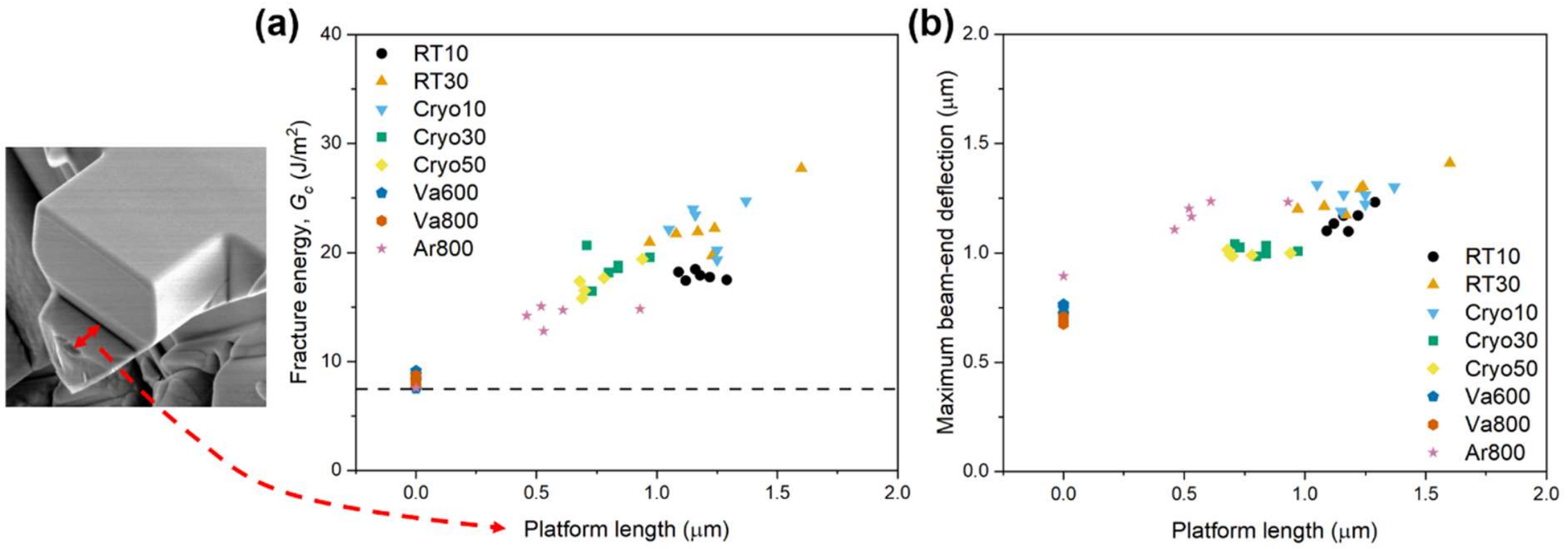

**Figure 6**. (a) Fracture energy, $G_c$, as a function of platform length for the eight sample groups. (b) Maximum beam-end deflection as a function of platform length for all sample groups. The dashed horizontal line in (a) indicates the $G_c$ value from DCB testing (7.5 J/m$^2$).

# 4. Discussion

## 4.1. Geometry-dependent reliability of microscale fracture properties

The fracture energy obtained from DCB splitting ($G_c$ = 7.5 ± 0.3 J/m$^2$) closely matches the modelled 6H-SiC $\{10\bar{1}0\}$ surface energy (2γ = 7.48 J/m$^2$) obtained from self-consistent charge density-functional based tight-binding (SCC-DFTB) method [48]. The value is also within the range reported for the prismatic $\{10\bar{1}0\}$ plane of 6H-SiC obtained via indentation (4.2 to 8.1 J/m$^2$) [49]. For comparison with the basal {0001} plane, Sernicola et al. reported 6 ± 2 J/m$^2$ from microscale DCB experiments [29]. Density functional theory (DFT) calculations by Remakers et al. yielded basal-plane fracture energies (2γ) ranging from 4.86 to 7.82 J/m$^2$, depending on the specific surface termination [50]. The present value for the prismatic plane lies near the upper end of the established energetic window for the basal plane in 6H-SiC. This is consistent with the observation that the basal plane is the most easily fractured cleavage plane in hexagonal SiC [51, 52].

A key finding of this study is the large discrepancy between the DCB-derived prismatic-plane fracture energy ($G_c$ = 7.5 ± 0.3 J/m$^2$) and the fracture energy values obtained from SCBs with straight-through notches prepared using a Ga-FIB without any post-notching treatment ($G_c$ ~ 18 to 22 J/m$^2$ depending on notching currents). Elevated fracture energy values have been reported previously in SCB tests of $Al_2O_3$ and were attributed to compressive residual stresses introduced by Ga implantation during the final FIB notching step [35]. Increasing the final notching current from 10 pA to 30 pA produced a modest increase in the measured fracture energy, likely associated with a blunter notch root produced at higher ion beam currents and thus greater resistance to fracture [35]. The DCB geometry offers one important advantage in this context. Stable crack growth under displacement control enables the tracking of $G_c$ as a function of crack extension; this excludes the potentially inaccurate fracture energy values near the notch tip where the FIB damage is the most influential. On the other hand, in SCBs with straight-through notches, the fracture energy is extracted from a single catastrophic event and is therefore inherently more sensitive to the near-notch conditions. Accordingly, SCB-derived $G_c$ and $K_{Ic}$ values capture not only the intrinsic cleavage resistance of the material, but also the influence of FIB-induced modifications of the near-notch stress state.

It is important to bear in mind that **Figure 2(c)** also highlights a drawback of the DCB geometry. As discussed by Sernicola et al. and Gavalda-Diaz et al., fracture energy values obtained during the early stages of DCB crack growth are less reliable than those measured after further crack extension [28, 29]. At short crack lengths, the crack opening is small; thus, the beam deflection and crack-tip position are more difficult to determine accurately. Under these conditions, errors in locating the crack tip can significantly affect the calculated fracture energy because of its strong inverse dependence on the fourth power of the crack length (**Equation 2**). As the crack propagates further away from the notch, the crack tip position and beam displacement data can be measured and fitted more reliably, so the calculated fracture energy values converge toward a stable plateau. This, compounded with the effect of FIB damage (typically tens of nanometres from the surface [33, 53-55]) on the measured fracture energy, likely led to the elevated values obtained during the initial 2-3 μm of crack growth in **Figure 2(c)**. As a consequence of this limitation of DCB testing, adopting this geometry may be challenging for materials in which sustaining long, straight crack propagation is difficult (e.g., a short grain boundary).

### 4.2. Effectiveness and limitations of near-cryogenic notching in SCBs

Previous cryo-FIB work has mainly demonstrated improved preservation of microstructural features or limitation of the redistribution of implanted species [39, 40]. The present results extend this concept to a more mechanically sensitive situation. In this case, the key question is not only whether cryo-FIB reduces preparation artefacts, but whether it sufficiently help recover intrinsic fracture properties. The comparison with the DCB benchmark and reported theoretical values shows that near-cryogenic notching alone is insufficient for this purpose. The low-temperature-notched specimens still exhibited higher fracture energies than those obtained from DCB measurements. The higher $G$ measured for the 10 pA notched SCBs, as compared to the ones notched with 30 and 50 pA FIB, may be attributed to the drift of the sample during low-temperature notching. Although thermal drift was minimised by allowing extended stabilisation period before milling, factors such as the continuous fluctuations in $N_2$ gas flow required to cool the stage still produced subtle drift with rates of up to ~20 nm/min, as can be observed in **Figure S1**. When 10 pA beam current was used, the final notching step required approximately 120 s (**Table 1**). Drift during this period may still alter the final notch root geometry (e.g., tip radius and symmetry), which can then bias the measured fracture properties. In contrast, at 30 pA and 50 pA, the notch was completed much faster, within 43 s and 29 s, respectively, reducing the potential for drift-induced notch-shape distortion.

Notably, under near-cryogenic conditions, increasing the final notching current from 10 pA to 30 pA and 50 pA led to a decrease in the measured fracture energies, in contrast to the room-temperature notching behaviour where higher current led to higher fracture energies. Room-temperature FIB milling is typically more stable because the stage-holder-sample assembly operates near thermal equilibrium, without active cooling that drives time-dependent drift under low-temperature conditions [56]. Therefore, at room temperature, the blunter notch tip produced at higher FIB current, rather than drift-induced notch distortion, becomes the dominant factor affecting the measured fracture energy, giving rise to the opposite trends observed at room-temperature and near-cryogenic conditions.

The measured fracture energies for both groups notched at the near-cryogenic temperature, Cryo30 and Cryo50, were lower than those obtained for the room-temperature notched group RT30. Despite the potential benefits of low-temperature milling, the fracture energy values were still higher than the DCB values. As discussed earlier, the principal desired advantage of cryo-FIB is the suppression of thermally activated redistribution of implanted ions and defects after the initial FIB-material interaction. Nonetheless, the primary implantation depth is governed predominantly by the ion energy and the angle of incidence [57], rather than the specimen temperature which is typically not explicitly considered in, for instance, stopping and range of ions in matter (SRIM) simulations [58]. Consequently, because the accelerating voltage and milling geometry were identical for room-temperature and low-temperature notching in this study, the initial implanted/damaged layer is not expected to differ substantially between the two conditions, aside from any damage recovery that may occur at room temperature. Under these circumstances, suppressing post-implantation redistribution at the near-cryogenic temperature alone is unlikely to fully mitigate the near-tip compressive residual stress field, which may explain why notching at the near-cryogenic temperature did not bring the SCB-derived fracture energies into agreement with the DCB geometry.

In summary, using near-cryogenic FIB notching to suppress irradiation-induced defects is insufficient for obtaining intrinsic fracture properties of the material, as the ion-implantation-induced compressive stress field remains to be a major contributor to the elevated apparent fracture properties measured.

### 4.3. Vacuum annealing to mitigate FIB-induced artefacts in SCB

As shown in **Section 3.2.3** and **Figure 5**, vacuum annealing at 600 °C resulted in the development of spherical Ga particles. A comparable phenomenon was reported by Jaya et al., who observed that, when Ga-implanted silicon was heated in vacuum, the implanted Ga migrated to and desorbed at the sample surface [59]. Once Ga accumulated at the surface, it tended to aggregate, consistent with the behaviour observed in the present study. Quantitatively, vacuum annealing at both 600 °C and 800 °C reduced the measured fracture energies to values close to the DCB results, accompanied by a substantial reduction in detectable Ga in the specimens (**Figure 5**). In agreement with previous studies, this result demonstrates that thermal treatment can largely remove FIB-preparation-induced residual stresses and the associated elevation in SCB-derived fracture energy [10, 35, 37], enabling SCB tests to more closely reflect the intrinsic fracture properties of materials.

A crucial finding here is that the annealing environment (vacuum vs. inert) has a pronounced effect on the results. Annealing at 800 °C and ~$10^5$ Pa in Ar produced only a moderate decrease in $G_c$, and widespread Ga particles were observed on the specimen surface (**Figure 4(c)**). This behaviour is consistent with the strong pressure dependence of Ga volatility: under vacuum ($10^{-3}$ to $10^{-4}$ Pa) the effective boiling temperature is substantially reduced from 2400 °C at atmospheric pressure to 755 °C to 681 °C, promoting Ga evaporation during annealing at 800 °C. Nonetheless, under ~$10^5$ Pa of Ar, Ga is far less volatile (boiling point ~2400 °C) [46], hence may melt and redistribute before resolidifying as particles. Although the reduced fracture energy for specimens annealed in Ar suggests some Ga redistribution away from the notch tip and potentially partial lattice recovery [60], the persistently elevated fracture energy measured for Ar800 relative to the vacuum-annealed specimens implies that Ga removal from the material may not be as thorough.

While these results demonstrate the effectiveness of vacuum annealing under the conditions tested, the present study did not identify the lowest annealing temperature or shortest annealing duration needed to mitigate FIB-induced artefacts. The annealing conditions used here were selected to promote Ga redistribution and removal below and above the estimated Ga boiling points (755 °C and 681 °C) under vacuum ($10^{-3}$ and $10^{-4}$ Pa). The effectiveness of the 600 °C vacuum anneal suggests that complete Ga evaporation is not necessarily required to remove the preparation-induced artefacts. This raises the possibility that lower-temperature annealing may also be effective.

### 4.4. Platform features in SCB

An additional observation of this work is the systematic association between the step-like platform feature and the elevated SCB fracture energies (**Figure 6**). FIB notching with Ga introduces a localised compressive residual stress field at the notch tip that can shield the crack tip [35], thereby increasing the applied load required to initiate cracking. The higher initiation load naturally allows greater beam-end deflection before fracture. As deflection increases, the stress state at the crack tip likely becomes more complex and deviates from ideal Mode I, potentially resulting in mixed-mode loading, which may promote crack deflection when a shear component is present [61, 62]. In 6H-SiC, cracking parallel to the basal plane is easier than in other orientations [51, 52]. Thus, a small off-axis stress component at the crack tip may promote limited cracking along the basal plane as observed in **Figure 3(c)**.

### 4.5. Alternative SCB notch geometries

Reliable fracture-energy measurements require minimising the influence of the FIB-modified notch region. In this study, this was achieved either by using DCBs, where data near the initial notch could be discarded, or by post-notching annealing of SCBs. However, DCB geometry may not always be feasible because of geometric constraints, and annealing may alter the material state. Alternative SCB notch geometries therefore offer another route to reduce FIB-induced artefacts [63-67]. Chevron notches, for example, reduce the influence of the FIB-modified notch root by promoting the formation of a sharp crack and stabilising its early growth before catastrophic failure [64, 65]. In brittle materials such as single-crystal silicon, this crack geometry can enable crack extension followed by arrest prior to unstable fracture, although crack initiation at the chevron tip may require pre-fatigue before loading [64].

Bridge-notched SCBs seek to reduce FIB-notch effects through a different mechanism. Thin material bridges are retained at the notch ends and are designed to fail before the main notch front. If the stress intensity at the main notch remains below the critical value after bridge failure, the cracks initiated at the bridges arrest, providing sharp, naturally formed cracks for the final fracture event [63, 66, 67]. This mechanism was first directly observed in CrN/AlN and CrN SCBs, where bridge failure produced load drops followed by arrested cracks prior to final fracture [63]. However, crack arrest is strongly geometry-dependent. In single-crystal silicon, it occurs only for sufficiently thin bridges and deep notches. Thicker bridges or shallower notches can suppress crack arrest and lead to apparent toughening [67]. Bridge-notched SCB data should therefore be interpreted as evidence of reduced FIB-notch effects only when both bridge failure and crack arrest are confirmed, for example through load drops in the load-displacement curve and direct *in situ* observation. Otherwise, it remains uncertain whether final fracture starts from the bridge-generated cracks or from the original FIB notch front [67].

## 5. Conclusion

This study provides a systematic comparison of microscale fracture energy measurements obtained from DCB and SCB geometries in single-crystal 6H-SiC, with particular emphasis on the influence of notch fabrication and post-processing in SCBs. Wedge-loaded DCB experiments exhibited stable crack growth and yielded a reproducible fracture energy of 7.5 ± 0.3 J/m$^2$ for cleavage along the prismatic $\{10\bar{1}0\}$ plane. The ability to track fracture energy as a function of crack extension makes the DCB geometry inherently robust against near-notch artefacts introduced during FIB preparation.

SCBs with straight-through Ga-FIB-milled notches showed substantially elevated fracture energies when tested without post-processing, reflecting the strong influence of Ga implantation-induced compressive residual stresses around the notch tip. A key new finding is that near-cryogenic Ga-FIB notching, while attractive as a damage-mitigation strategy, is not sufficient by itself to obtain intrinsic fracture properties from SCB measurements in 6H-SiC. Post-notching annealing, particularly under vacuum, proved to be the most effective strategy for reducing preparation-induced artefacts in SCBs. Vacuum annealing at 600 °C and 800 °C significantly reduced detectable Ga content and produced fracture energy values comparable to those measured using DCBs, while annealing in an Ar atmosphere was less effective.

Overall, this work demonstrates that microscale fracture measurements are highly sensitive to both testing geometry and specimen preparation. DCB testing offers advantages in data reliability, although the experimental process can be practically demanding (e.g., precise tip-notch alignment). In contrast, SCB geometries offer greater testing flexibility (e.g., possible using various tip types, easier alignment), but the measured fracture properties are strongly affected by Ga implantation and residual stresses at the notch root. Reliable SCB-based measurements therefore require careful mitigation of FIB-induced damage, with vacuum annealing as an effective approach. Together, these findings provide guidelines for selecting appropriate test geometries and preparation routes, and help improve the accuracy and reproducibility, of small-scale fracture measurements in ceramics and other brittle materials.

## Acknowledgements

This work was supported by the EPSRC InFUSE Prosperity Partnership (EP/V038044/1). The authors acknowledge the Harvey Flowers Electron Microscopy Suite and Henry Royce Institute (Royce at Imperial) for access to facilities and thank Dr. Mahmoud Ardakani and Dr. Ecaterina (Cati) Ware for training on the equipment. SW acknowledges Imperial College London for funding his Imperial College Research Fellowship.

## References

[1] V. Jayaram, Small-scale mechanical testing, Annual Review of Materials Research 52(1) (2022) 473-523.

[2] M. Emmanuel, O. Gavalda-Diaz, G. Sernicola, R. M'saoubi, T. Persson, S. Norgren, K. Marquardt, T.B. Britton, F. Giuliani, Fracture energy measurement of prismatic plane and Σ2 boundary in cemented carbide, JOM 73(6) (2021) 1589-1596.

[3] D. Casellas, J. Caro, S. Molas, J.M. Prado, I. Valls, Fracture toughness of carbides in tool steels evaluated by nanoindentation, Acta Materialia 55(13) (2007) 4277-4286.

[4] K.I. Schiffmann, Determination of fracture toughness of bulk materials and thin films by nanoindentation: comparison of different models, Philosophical Magazine 91(7-9) (2011) 1163-1178.

[5] G. Pharr, D. Harding, W. Oliver, Measurement of fracture toughness in thin films and small volumes using nanoindentation methods, Mechanical properties and deformation behavior of materials having ultra-fine microstructures, Springer1993, pp. 449-461.

[6] D. Harding, W. Oliver, G. Pharr, Cracking during nanoindentation and its use in the measurement of fracture toughness, MRS Online Proceedings Library (OPL) 356 (1994) 663.

[7] M. Sebastiani, K.E. Johanns, E.G. Herbert, G.M. Pharr, Measurement of fracture toughness by nanoindentation methods: Recent advances and future challenges, Current Opinion in Solid State and Materials Science 19(6) (2015) 324-333.

[8] C.M. Lauener, L. Petho, M. Chen, Y. Xiao, J. Michler, J.M. Wheeler, Fracture of Silicon: Influence of rate, positioning accuracy, FIB machining, and elevated temperatures on toughness measured by pillar indentation splitting, Materials & Design 142 (2018) 340-349.

[9] M.Z. Mughal, H.-Y. Amanieu, R. Moscatelli, M. Sebastiani, A Comparison of Microscale Techniques for Determining Fracture Toughness of LiMn2O4 Particles, Materials 10(4) (2017) 403.
[10] J.P. Best, J. Zechner, J.M. Wheeler, R. Schoeppner, M. Morstein, J. Michler, Small-scale fracture toughness of ceramic thin films: the effects of specimen geometry, ion beam notching and high temperature on chromium nitride toughness evaluation, Philosophical Magazine 96(32-34) (2016) 3552-3569.
[11] J. Ast, M. Ghidelli, K. Durst, M. Göken, M. Sebastiani, A.M. Korsunsky, A review of experimental approaches to fracture toughness evaluation at the micro-scale, Materials & Design 173 (2019) 107762.
[12] M. Sebastiani, K.E. Johanns, E.G. Herbert, F. Carassiti, G.M. Pharr, A novel pillar indentation splitting test for measuring fracture toughness of thin ceramic coatings, Philosophical Magazine 95(16-18) (2015) 1928-1944.
[13] B.N. Jaya, C. Kirchlechner, G. Dehm, Can microscale fracture tests provide reliable fracture toughness values? A case study in silicon, Journal of Materials Research 30(5) (2015) 686-698.
[14] B.N. Jaya, S. Bhowmick, S.A.S. Asif, O.L. Warren, V. Jayaram, Optimization of clamped beam geometry for fracture toughness testing of micron-scale samples, Philosophical Magazine 95(16-18) (2015) 1945-1966.
[15] K. Venkatraman, V. Jayaram, Stiffness based technique to probe cyclic damage accumulation in micro-structurally graded bond coats via micro-beam bending tests, Philosophical Magazine 99(16) (2019) 2016-2050.
[16] N. Jaya B, V. Jayaram, S.K. Biswas, A new method for fracture toughness determination of graded (Pt,Ni)Al bond coats by microbeam bend tests, Philosophical Magazine 92(25-27) (2012) 3326-3345.
[17] J. Jiang, S. Falco, S. Wang, F. Giuliani, R.I. Todd, Microcantilever investigation of slow crack growth and crack healing in aluminium oxide, Acta Materialia 273 (2024) 119914.
[18] J. Tatami, M. Katayama, M. Ohnishi, T. Yahagi, T. Takahashi, T. Horiuchi, M. Yokouchi, K. Yasuda, D.K. Kim, T. Wakihara, K. Komeya, Local Fracture Toughness of Si3N4 Ceramics Measured using Single-Edge Notched Microcantilever Beam Specimens, Journal of the American Ceramic Society 98(3) (2015) 965-971.
[19] D.E.J. Armstrong, A.S.M.A. Haseeb, S.G. Roberts, A.J. Wilkinson, K. Bade, Nanoindentation and micro-mechanical fracture toughness of electrodeposited nanocrystalline Ni–W alloy films, Thin Solid Films 520(13) (2012) 4369-4372.
[20] D.E.J. Armstrong, A.J. Wilkinson, S.G. Roberts, Micro-mechanical measurements of fracture toughness of bismuth embrittled copper grain boundaries, Philosophical Magazine Letters 91(6) (2011) 394-400.
[21] D.E.J. Armstrong, A.J. Wilkinson, S.G. Roberts, Measuring Local Mechanical Properties using FIB Machined Microcantilevers, MRS Online Proceedings Library 1185(1) (2009) 13-19.
[22] B.R. Mansfield, D.E.J. Armstrong, P.R. Wilshaw, J.D. Murphy, An Investigation into Fracture of Multi-Crystalline Silicon, Solid State Phenomena 156-158 (2010) 55-60.
[23] D. Armstrong, A. Haseeb, A. Wilkinson, S. Roberts, Micro-Fracture testing of Ni-W Microbeams Produced by Electrodeposition and FIB Machining, MRS Proceedings 983 (2006) 0983-LL08-07.
[24] S.M. Dickens, F.W. DelRio, S.J. Grutzik, W.M. Mook, B.L. Boyce, E. Hintsala, D. Stauffer, R.F. Cook, In-situ Fracture Toughness of Single Crystal Silicon Double-Cantilever Beams, Microscopy and Microanalysis 28(S1) (2022) 14-15.
[25] Y. Piao, S. Wang, D.S. Balint, F. Giuliani, E. Saiz, O. Gavalda-Diaz, Deformation and toughening of graphite at the micron scale, Acta Materialia 299 (2025) 121428.
[26] F.W. DelRio, S.J. Grutzik, W.M. Mook, S.M. Dickens, P.G. Kotula, E.D. Hintsala, D.D. Stauffer, B.L. Boyce, Eliciting stable nanoscale fracture in single-crystal silicon, Materials Research Letters 10(11) (2022) 728-735.

[27] O. Gavalda-Diaz, J. Lyons, S. Wang, M. Emmanuel, K. Marquardt, E. Saiz, F. Giuliani, Basal plane delamination energy measurement in a Ti3SiC2 MAX phase, Jom 73(6) (2021) 1582-1588.

[28] O. Gavalda-Diaz, R. Manno, A. Melro, G. Allegri, S.R. Hallett, L. Vandeperre, E. Saiz, F. Giuliani, Mode I and Mode II interfacial fracture energy of SiC/BN/SiC CMCs, Acta Materialia 215 (2021) 117125.

[29] G. Sernicola, T. Giovannini, P. Patel, J.R. Kermode, D.S. Balint, T.B. Britton, F. Giuliani, In situ stable crack growth at the micron scale, Nature communications 8(1) (2017) 108.

[30] D. Di Maio, S. Roberts, Measuring fracture toughness of coatings using focused-ion-beam-machined microbeams, Journal of materials research 20(2) (2005) 299-302.

[31] A. Stratulat, D.E. Armstrong, S.G. Roberts, Micro-mechanical measurement of fracture behaviour of individual grain boundaries in Ni alloy 600 exposed to a pressurized water reactor environment, Corrosion Science 104 (2016) 9-16.

[32] S. Liu, J. Wheeler, P. Howie, X. Zeng, J. Michler, W. Clegg, Measuring the fracture resistance of hard coatings, Applied Physics Letters 102(17) (2013).

[33] L. Borasi, A. Slagter, A. Mortensen, C. Kirchlechner, On the preparation and mechanical testing of nano to micron-scale specimens, Acta Materialia 283 (2025) 120394.

[34] Y. Chen, X. Zhang, X. Zhao, N. Markocsan, P. Nylén, P. Xiao, Measurements of elastic modulus and fracture toughness of an air plasma sprayed thermal barrier coating using micro-cantilever bending, Surface and Coatings Technology 374 (2019) 12-20.

[35] A. Norton, S. Falco, N. Young, J. Severs, R. Todd, Microcantilever investigation of fracture toughness and subcritical crack growth on the scale of the microstructure in Al2O3, Journal of the European Ceramic Society 35(16) (2015) 4521-4533.

[36] J.P. Best, J. Zechner, I. Shorubalko, J.V. Oboňa, J. Wehrs, M. Morstein, J. Michler, A comparison of three different notching ions for small-scale fracture toughness measurement, Scripta Materialia 112 (2016) 71-74.

[37] E. Okotete, S. Mueck, S. Lee, C. Kirchlechner, Evaluating neon ions as an alternative to gallium in micro cantilevers fracture testing, Scripta Materialia 258 (2025) 116509.

[38] R. Estivill, G. Audoit, J.-P. Barnes, A. Grenier, D. Blavette, Preparation and Analysis of Atom Probe Tips by Xenon Focused Ion Beam Milling, Microscopy and Microanalysis 22(3) (2016) 576-582.

[39] Y. Chang, W. Lu, J. Guénolé, L.T. Stephenson, A. Szczpaniak, P. Kontis, A.K. Ackerman, F.F. Dear, I. Mouton, X. Zhong, S. Zhang, D. Dye, C.H. Liebscher, D. Ponge, S. Korte-Kerzel, D. Raabe, B. Gault, Ti and its alloys as examples of cryogenic focused ion beam milling of environmentally-sensitive materials, Nature Communications 10(1) (2019) 942.

[40] L. Lilensten, B. Gault, New approach for FIB-preparation of atom probe specimens for aluminum alloys, PLOS ONE 15(4) (2020) e0231179.

[41] J. Williams, End corrections for orthotropic DCB specimens, Composites Science and Technology 35(4) (1989) 367-376.

[42] S. Wang, O. Gavalda-Diaz, J. Lyons, F. Giuliani, Shear and delamination behaviour of basal planes in Zr3AlC2 MAX phase studied by micromechanical testing, Scripta Materialia 240 (2024) 115829.

[43] N. Aldegaither, G. Sernicola, A. Mesgarnejad, A. Karma, D. Balint, J. Wang, E. Saiz, S.J. Shefelbine, A.E. Porter, F. Giuliani, Fracture toughness of bone at the microscale, Acta Biomaterialia 121 (2021) 475-483.

[44] K. Kamitani, M. Grimsditch, J. Nipko, C.-K. Loong, M. Okada, I. Kimura, The elastic constants of silicon carbide: A Brillouin-scattering study of 4H and 6H SiC single crystals, Journal of applied physics 82(6) (1997) 3152-3154.

[45] F. Iqbal, J. Ast, M. Göken, K. Durst, In situ micro-cantilever tests to study fracture properties of NiAl single crystals, Acta Materialia 60(3) (2012) 1193-1200.

[46] Y. Zhang, J.R.G. Evans, S. Yang, Corrected Values for Boiling Points and Enthalpies of Vaporization of Elements in Handbooks, Journal of Chemical & Engineering Data 56(2) (2011) 328-337.

[47] S. Velasco, F. Román, J. White, On the Clausius–Clapeyron vapor pressure equation, Journal of Chemical Education 86(1) (2009) 106.

[48] E. Rauls, Z. Hajnal, P. Deak, T. Frauenheim, Theoretical study of the nonpolar surfaces and their oxygen passivation in 4 H-and 6 H-SiC, Physical Review B 64(24) (2001) 245323.

[49] H. Kitahara, Y. Noda, F. Yoshida, H. Nakashima, N. Shinohara, H. Abe, Mechanical behavior of single crystalline and polycrystalline silicon carbides evaluated by Vickers indentation, Journal of the Ceramic Society of Japan 109(1271) (2001) 602-606.

[50] S. Ramakers, A. Marusczyk, M. Amsler, T. Eckl, M. Mrovec, T. Hammerschmidt, R. Drautz, Effects of thermal, elastic, and surface properties on the stability of SiC polytypes, Physical Review B 106(7) (2022) 075201.

[51] K. Kishida, Y. Shinkai, H. Inui, Room temperature deformation of 6H–SiC single crystals investigated by micropillar compression, Acta Materialia 187 (2020) 19-28.

[52] B. Meng, C. Li, Effect of anisotropy on deformation and crack formation under the brittle removal of 6H-SiC during SPDT process, Journal of Advanced Research 56 (2024) 103-112.

[53] M. Presley, J. Jensen, D. Huber, H. Fraser, Exploring Differences in Amorphous Layer Formation during FIB Sample Preparation between Metals and Non Metals, Microscopy and Microanalysis 22(S3) (2016) 144-145.

[54] S. Rubanov, P.R. Munroe, FIB-induced damage in silicon, Journal of Microscopy 214(3) (2004) 213-221.

[55] J. Mayer, L.A. Giannuzzi, T. Kamino, J. Michael, TEM Sample Preparation and FIB-Induced Damage, MRS Bulletin 32(5) (2007) 400-407.

[56] V. Lam, E. Villa, Practical Approaches for Cryo-FIB Milling and Applications for Cellular Cryo-Electron Tomography, in: T. Gonen, B.L. Nannenga (Eds.), cryoEM: Methods and Protocols, Springer US, New York, NY, 2021, pp. 49-82.

[57] R. Steve, P. Robert, A review of focused ion beam applications in microsystem technology, Journal of Micromechanics and Microengineering 11(4) (2001) 287.

[58] J.F. Ziegler, J.P. Biersack, The Stopping and Range of Ions in Matter, in: D.A. Bromley (Ed.), Treatise on Heavy-Ion Science: Volume 6: Astrophysics, Chemistry, and Condensed Matter, Springer US, Boston, MA, 1985, pp. 93-129.

[59] B.N. Jaya, J.M. Wheeler, J. Wehrs, J.P. Best, R. Soler, J. Michler, C. Kirchlechner, G. Dehm, Microscale Fracture Behavior of Single Crystal Silicon Beams at Elevated Temperatures, Nano Letters 16(12) (2016) 7597-7603.

[60] J.-J. Gu, J.-H. Zhao, M.-Y. Bu, S.-M. Wang, L. Fan, Q. Huang, S. Li, Q.-Y. Yue, X.-L. Wang, Z.-X. Wei, Y. Liu, Study on the damage evolution of 6H-SiC under different phosphorus ion implantation conditions and annealing temperatures, Results in Physics 43 (2022) 106127.

[61] F. Erdogan, G.C. Sih, On the Crack Extension in Plates Under Plane Loading and Transverse Shear, Journal of Basic Engineering 85(4) (1963) 519-525.
[62] B. Cotterell, J.R. Rice, Slightly curved or kinked cracks, International Journal of Fracture 16(2) (1980) 155-169.
[63] Y. Zhang, M. Bartosik, S. Brinckmann, S. Lee, C. Kirchlechner, Direct observation of crack arrest after bridge notch failure: A strategy to increase statistics and reduce FIB-artifacts in micro-cantilever testing, Materials & Design 233 (2023) 112188.
[64] M.G. Mueller, G. Žagar, A. Mortensen, Stable room-temperature micron-scale crack growth in single-crystalline silicon, Journal of Materials Research 32(19) (2017) 3617-3626.
[65] B.S. Li, T.J. Marrow, S.G. Roberts, D.E.J. Armstrong, Evaluation of Fracture Toughness Measurements Using Chevron-Notched Silicon and Tungsten Microcantilevers, JOM 71(10) (2019) 3378-3389.
[66] N.G. Mathews, A.K. Mishra, B.N. Jaya, Mode dependent evaluation of fracture behaviour using cantilever bending, Theoretical and Applied Fracture Mechanics 115 (2021) 103069.
[67] E. Okotete, A. Muslija, J.K. Hohmann, M. Kohl, S. Brinckmann, S. Lee, C. Kirchlechner, Enhanced crack stability in micro scale fracture testing via optimized bridge notches, Materials Science and Engineering: A 939 (2025) 148479.

# Supplementary information: Fracture energy of 6H-SiC at the microscale: effects of testing geometry and notch preparation

*Zhuoqi Lucas Li [a,*], Siyang Wang [a], Ao Li [a], James O. Douglas [a], Florian Bouville [a], Oriol Gavalda-Diaz [a], Katharina Tinka Marquardt [a,b,c], Finn Giuliani [a]*

[a] *Department of Materials, Imperial College London, London SW7 2AZ, UK*

[b] *Department of Materials, University of Oxford, Oxford OX2 6NN, UK*

[c] *The Zero Institute, Holywell House, Osney Mead, Oxford OX2 0ES, UK*

*Corresponding author: Zhuoqi Lucas Li (zhuoqi.li17@imperial.ac.uk)*

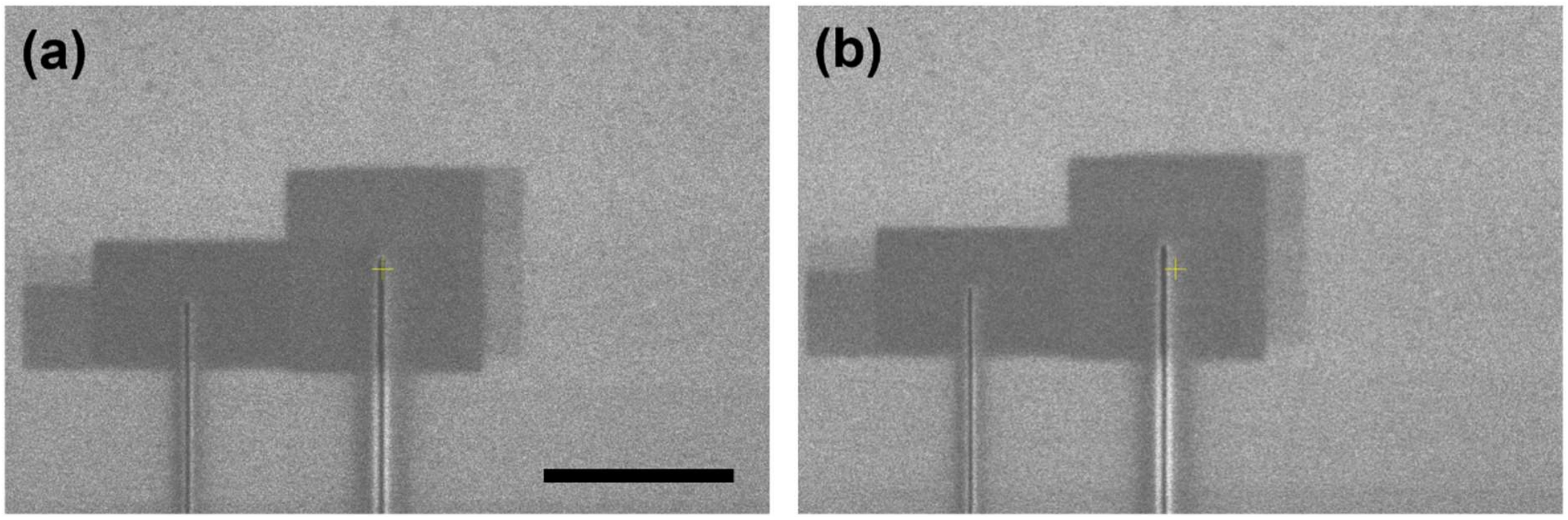

**Figure S1.** Drift during near-cryogenic FIB milling at (a) time zero and (b) after 5 min. The centre crosshair shifted during the observation period, corresponding to horizontal and vertical drift rates of ~20 nm/min. Scale bars: 2 μm.